\documentclass[preprintnumbers,amsmath,amssymb,prl]{revtex4-1}
\usepackage{graphicx}    
\usepackage{epsfig}      
\usepackage{dcolumn}     
\usepackage{bm}          
\usepackage{color}
\newcounter{saveenumi}

\newcommand{\be}{\begin{enumerate}}
\newcommand{\ee}{\end{enumerate}}

\newcommand{\RAA}{\AA$^{-1}$}
\def\cis{CuIr$_2$S$_4$}
\def\cise{CuIr$_2$Se$_4$}
\def\cics{Cu(Ir$_{1-x}$Cr$_x$)$_2$S$_4$}
\def\czis{(Cu$_{1-x}$Zn$_x$)Ir$_2$S$_4$}

\def\ranget{10~K $\le$ T $\le$ 780~K}

\def\fd3m{Fd$\overline 3$m}
\def\p1bar{P$\overline 1$}
\def\i41amd{I4$_{1}$/amd}
\def\t2g{$t_{2g}$}
\def\tmit{T$_{MI}$}
\begin{document}
%
%
\begin{abstract}
{\bf Fundamental electronic principles underlying all transition metal compounds are the symmetry and filling of the $d$-electron orbitals and the influence of this filling on structural configurations and responses. Here we use a sensitive local structural technique, x-ray atomic pair distribution function analysis, to reveal the presence of fluctuating local-structural distortions at high temperature of one such compound, \cis . We show that this hitherto overlooked fluctuating symmetry lowering is electronic in origin and will significantly modify the energy-level spectrum and electronic and magnetic properties. The explanation is a local, fluctuating,  orbital-degeneracy-lifted state.  The natural extension of our result would be that this phenomenon is likely to be widespread amongst diverse classes of partially filled nominally degenerate d-electron systems, with potentially broad implications for our understanding of their properties.}
\end{abstract}
\title{Local orbital degeneracy lifting as a precursor to an orbital-selective Peierls transition}

    \author{E.~S.~Bozin$^{1,*}$, W.~G.~Yin$^{1}$, R.~J.~Koch$^{1}$, M.~Abeykoon$^{2}$, Y.~S.~Hor$^{3,\dag}$, H.~Zheng$^3$, H.~C. Lei$^{1,\dag\dag}$, C.~Petrovic$^{1}$, J.~F.~Mitchell$^{3}$, and S.~J.~L.~Billinge$^{1,4}$}

    \affiliation{$^{1}$Condensed Matter Physics and Materials Science Department, Brookhaven National Laboratory, Upton, NY~11973, USA}
    \altaffiliation{bozin@bnl.gov\\
                    $^\dag$ Present address: Department of Physics, Missouri University of Science and Technology, Rolla, MO 65409, USA\\
                    $^{\dag\dag}$Present address: Department of Physics and Beijing Key Laboratory of Opto-electronic Functional Materials and Micro-nano Devices, Renmin University of China, Beijing 100872, China.}
    \affiliation{$^{2}$Photon Sciences Division, Brookhaven National Laboratory, Upton, NY 11973, USA}
    \affiliation{$^{3}$Materials Science Division, Argonne National Laboratory, Argonne, Illinois 60439, USA}
    \affiliation{$^{4}$Department of Applied Physics and Applied Mathematics, Columbia University, New York, NY~10027, USA}

\date{\today}
\maketitle
%
%
Broken symmetry ground states are often found in transition metal systems exhibiting emergent properties such as metal-insulator transitions~\cite{aetuk;np13,tian;np16,liang;np17}, charge ordered and charge density wave states~\cite{achka;nm16}, colossal magnetoresistive effects~\cite{savit;nc17}, frustrated magnetism~\cite{zorko;nc14,glasb;np15}, pseudogap~\cite{boris;prl08} and high temperature superconductivity~\cite{keime;n15,wang;np15}.
These are generally driven by electronic interactions understood as Fermi-surface nesting~\cite{chuan;s01,johan;prb08,teras;pnas09}, Peierls distortions~\cite{lee;prl73,bhobe;prx15} and cooperative Jahn-Teller effects~\cite{huang;nc17}. These phenomena have energy scales of hundreds to thousands of meV~\cite{khoms;b;tmc14}, corresponding to thousands of degrees kelvin, yet the broken symmetries tend to appear at much lower temperatures, typically 10$^1$-10$^2$~K.
The symmetry broken states at low temperature, have been extensively studied. Fewer details are known about what happens when these materials transition to crystallographically higher symmetry structures upon warming.

Here we present a study that reveals critical insights into the unaccounted for separation in energy scales by applying a quantitative local structure probe, atomic pair distribution function analysis (PDF), to a model material system that exemplifies this behavior. The material system, \cis, has rich broken symmetries in its ground-state~\cite{radae;n02}, including the formation of magnetic singlet Ir-Ir pairs, which disappear on warming through a structural transition that is also, concurrently, a metal-insulator transition (MIT).  The PDF analysis reveals difficult to detect but important local structural distortions that exist at high temperatures, something that has been seen before in other systems (see for example~\cite{billi;prl96}).  However, the special scattering characteristics of this system, together with our detailed temperature and doping dependent study with multiple dopant species, exposes fine details of the phenomenon establishing it as a robust but fluctuating $d$-orbital-degeneracy-lifted (ODL) state that is observed to the highest temperatures measured.  Much lower elastic energies govern the long-range ordering of the pre-formed local symmetry broken ODL objects in these structurally compliant materials~\cite{billi;prb02} which therefore occurs at temperatures much lower than the electronic energies of ODL formation. Interestingly, in \cis\ it is not the formation of the ODL objects, but their ordering that precipitates the MIT and magnetic dimer formation of the ground-state. The fluctuating ODL state is stabilized electronically by breaking $d$-electron orbital degeneracies and as such is likely to be a phenomenon that is widespread, though not widely appreciated, among the many materials with incompletely filled $d$-electron manifolds~\cite{sprau;s17,kosti;nm18}, many of which have important emergent low temperature electronic and magnetic behaviors, from classics such as manganites~\cite{chuan;s01,masse;np11,panop;njpqm18}, cuprates~\cite{achka;nm16,keime;n15,scagn;s11}, and iron chalcogenides/pnictides~\cite{glasb;np15,wang;np15,wen;prl12,kasah;n12,baek;nm15,frand;prb18}, to materials featuring exotic low-temperature orbital molecules~\cite{radae;njp05,attfi;aplm15}.  It may also explain the unexpected observation of phonon-glass-like thermal conductivity in various transition metal oxides~\cite{rivas;prb11}

The low temperature insulating state in \cis~\cite{furub;jpsj94,matsu;prb97,nagat;prb98} consists of ordered Ir$^{3+}$ ($5d^{6}$) and Ir$^{4+}$ ($5d^{5}$) ions~\cite{takub;prl05}, with a four-fold periodicity, an example of tetrameric charge ordering~\cite{croft;prb03}.
Concurrently, spin dimerization of Ir$^{4+}$ pairs occurs within the tetramer, with large associated structural distortions as they move towards each other, making this charge-order particularly amenable to detection using structural probes~\cite{radae;n02}.
Notwithstanding the complexities of the insulating state, including formation of remarkable three-dimensional Ir$^{3+}_{8}$S$_{24}$ and Ir$^{4+}_{8}$S$_{24}$ molecule-like assemblies embedded in the lattice, its quasi-one-dimensional character was unmasked and MIT attributed to an orbital-selective Peierls mechanism, postulated from topological considerations~\cite{khoms;prl05}. The global symmetry lowering at the MIT was declared to lift the existing \t2g\ $d$-orbital degeneracies~\cite{khoms;prl05}.
While the high temperature crystallographically cubic metallic state~\cite{furub;jpsj94,radae;n02} appears to be undistorted, with nominally Ir$^{3.5+}$ ($5d^{5.5}$) partially filled delocalized bands~\cite{bozin;prl11}, \cis\ does not behave like a band-metal, as evidenced by anomalous transport and spectroscopic signatures~\cite{burko;prb00,takub;prb08}.
Despite early speculations to the contrary~\cite{burko;prb00,yagas;jpsj06,takub;prb08}, it was established that the structural dimers disappear on warming through the transition on all lengthscales,
leaving the mystery of poor metallicity unresolved~\cite{bozin;prl11}.
Curiously, the isostructural and isoelectronic sister compound, \cise, has an order of magnitude higher conductivity and no MIT down to 0.5~K~\cite{burko;prb00}, which is also difficult to rationalize within the current understanding of these systems.

\begin{figure*}[tb]
\includegraphics[width=1.0\textwidth]{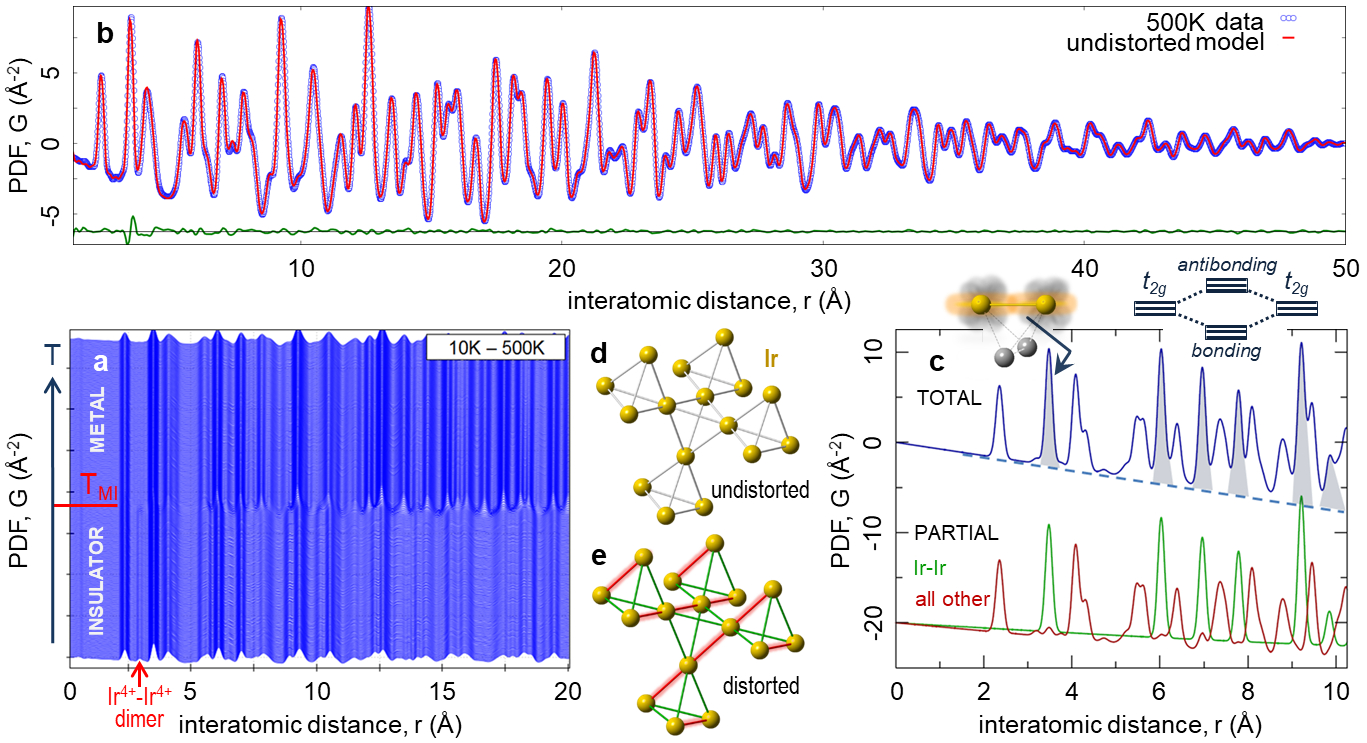}
\caption{\label{fig:onenm}{\bf Observation of high temperature fluctuating ODL state in \cis.} {\bf a}, Temperature waterfall stack of xPDFs measured on warming from 10~K (bottom) to 500~K (top) in 2~K increments. T$_{MI}$ is the MIT transition temperature (226~K). The dimer peak at $\sim$3~\AA\ (marked by arrow) is only seen in the insulating phase, and disappears above T$_{MI}$.
{\bf b}, Fit of the undistorted cubic \fd3m\ model (line) to the 500~K data (symbols) and their difference (offset for clarity) unmask the footprint of the localized ODL state at $\sim$3.5~\AA. {\bf c}, Simulated \fd3m\ total xPDF of \cis\ (blue line), decomposed into Ir-Ir partial xPDF (green line) and its complement (red line). Shaded peaks in total xPDF are sensitive to \t2g\ orbital overlaps (sketched) and their spatial correlations. Inset:  \t2g-derived molecular orbitals discussed in the main text. {\bf d},{\bf e}, Section of Ir pyrochlore sublattice of corner shared Ir$_{4}$ tetrahedra for undistorted (cubic) and distorted (tetragonal) spinel structure, respectively. The strongest \t2g\ orbital overlaps (e.g. {\em xy} with {\em xy}, etc.) are along the chains formed by the tetrahedral edges of the Ir sublattice.~\cite{khoms;prl05}}
\end{figure*}

In our high sensitivity x-ray pair distribution function (xPDF) analysis of the high-temperature metallic state of \cis\ we uncover a previously unobserved local symmetry lowering of the Ir pyrochlore sublattice, associated with an orbital liquid-like state that isi present the highest measured temperature. Through judicious chemical substitutions, we demonstrate that the effect is electronic and that it involves a symmetry lowering of the molecular orbitals, or Ir metal-metal bonds, on the pyrochlore sublattice.  This is related to, but qualitatively different from, the dimer-state observed in the insulating phase.
It is Jahn-Teller like, in that the symmetry lowering breaks the degeneracy of partially filled Ir $d$ states, which results in orbital selectivity, with charges preferrentially selecting a subset of Ir-Ir metal-metal orbitals.  At high temperature the selected orbitals do not order and are presumably fluctuating. This orbital liquid-like precursor state crystallizes upon approaching the Peierls-like MIT, testifying to the crucial role of orbital physics~\cite{khoms;prl05}.


%
%
\section{Structural fingerprint of the ODL state}
The PDF consists of peaks whose position is at interatomic distances in a material.
It is therefore sensitive to any structural perturbation, because sharp single-valued PDF peaks in a high symmetry structure become broadened or multicomponent when the symmetry is lowered.
In the low-temperature state of \cis , long range orbital and charge order results in Ir$^{4+}$-Ir$^{4+}$ pairs forming structural and magnetic dimers, which have been established crystallographically~\cite{radae;n02}.
The Ir-Ir dimer pair distance is about 0.5~\AA\ shorter than that of the Ir-Ir non-dimer pairs, creating two well-resolved peaks in the low temperature PDF.
In fact, the PDF dimer-peaks can be clearly seen by eye in the stack of PDFs shown as a function of temperature in Fig.~\ref{fig:onenm}(a) as a vertical ridge in the waterfall plot at $\approx 3$~\AA, labeled with the red arrow.

The dimers disappear in the average structure at $T_{MI}$~\cite{radae;n02}, but they also disappear in the local structure, as first reported in Ref.~\onlinecite{bozin;prl11}, and which can be seen directly in the data in Fig.~\ref{fig:onenm}(a).
There is no dimer-liquid state at high temperature in \cis, and the dimers themselves disappear at $T_{MI}$, which rules out fluctuating dimers as the culprits behind the poor metallicity at high-temperature.

We have approached the question of anomalous metallic state by measuring a new, more complete and higher precision set of xPDF data from \cis\ (Fig.~\ref{fig:onenm}(a)), where we now focus on the high temperature metallic state above $T_{MI}$.
The average crystal structure in this regime is cubic spinel, space-group \fd3m, in which
the iridium ions make a pyrochlore sublattice that consists of a network of regular corner-shared tetrahedra illustrated in Fig.~\ref{fig:onenm}(d).
The high symmetry of this cubic structure results in sharper peaks in the PDF, as is evident in the waterfall plot in Fig.~\ref{fig:onenm}(a), where PDF peak {\it sharpening} is observed {\it on warming} through \tmit\ (normally PDF peaks broaden on warming due to increased atomic thermal motion).
Indeed, fits of the cubic structure model to the high-T data result in excellent agreement (e.g., for $T=500$~K, r$_w$ =5.1\%, Fig.~\ref{fig:onenm}(b))
Under normal circumstances this would be considered a highly satisfactory PDF fit.
However, careful inspection of the residual curve in green in Fig.~\ref{fig:onenm}(b) reveals a single feature at around 3.5~\AA, indicating a shift in intensity to higher-$r$ in the data compared to the model.

The PDF peak centered at 3.5~\AA\ originates almost exclusively from the Ir-Ir nearest neighbor atomic pair on the pyrochlore sublattice, as shown in Fig.~\ref{fig:onenm}(c).
The total PDF consists of the weighted sum of partial PDFs between pairs of each type of atom, and Fig.~\ref{fig:onenm}(c) shows that the Ir-Ir partial-PDF contributes more than 95\% of the signal to the 3.5~\AA\ peak in the total PDF.
The residual signal therefore clearly originates from deviation of the structural geometry from the regular pyrochlore lattice implied by the cubic model.
Importantly, maximal \t2g\ overlaps of the orbitals of the same type ({\em xy} with {\em xy}, {\em yz} with {\em yz}, and {\em zx} with {\em zx}) are precisely along the directions defined by the edges of the pyrochlore lattice,~\cite{khoms;prl05} as sketched in the inset to Fig.~\ref{fig:onenm}(c), implying that the orbital sector is involved.

\section{Temperature evolution and characterization of the ODL state}
\begin{figure*}[tb]
\includegraphics[width=1.0\textwidth]{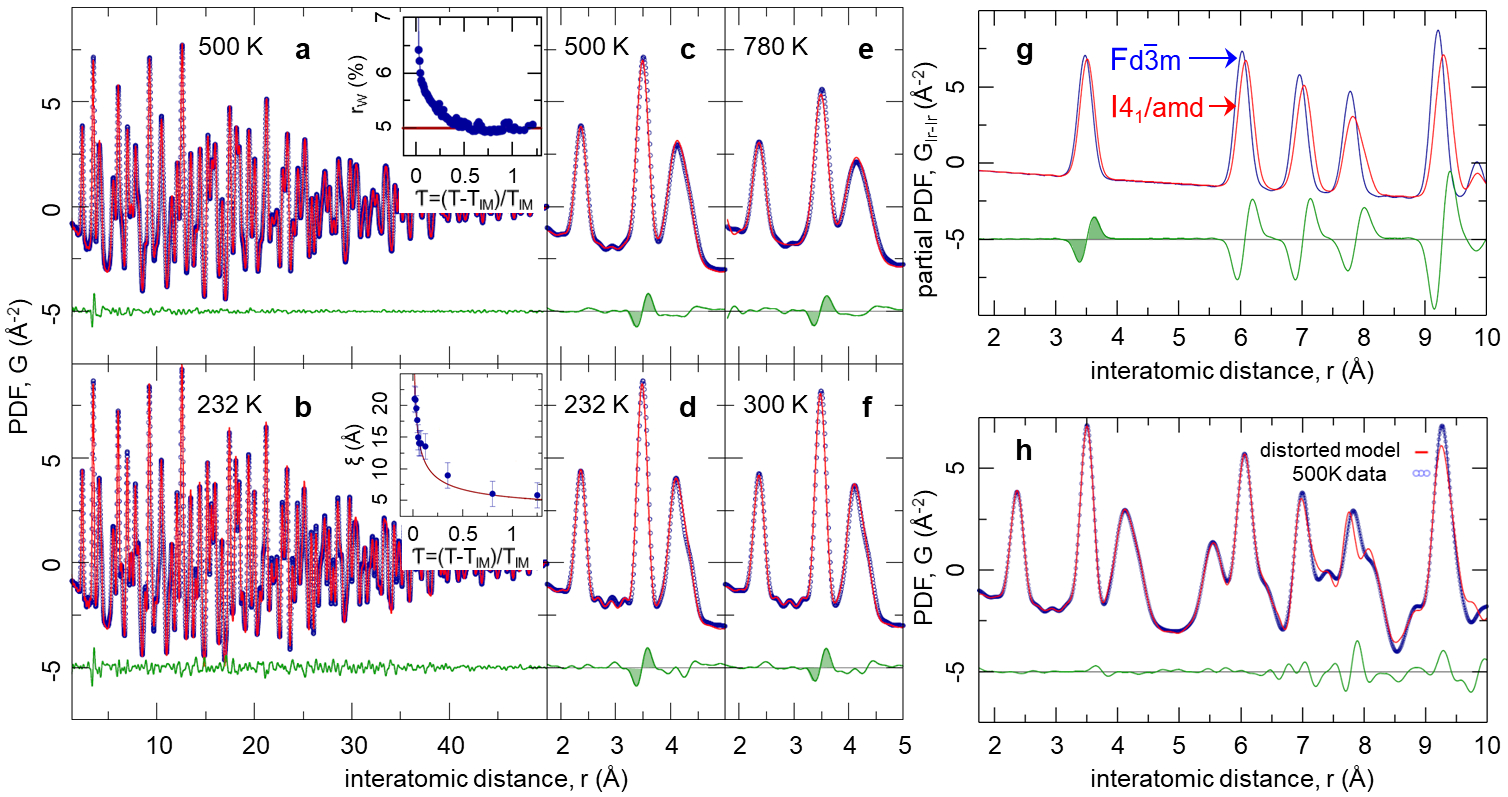}
\caption{\label{fig:twonm} {\bf Temperature evolution and character of the ODL state associated distortion.} {\bf a}-{\bf f}, Fit of the cubic \fd3m\ model (red line) to the \cis\ data at various temperatures as indicated (blue symbols). Difference curve between the data and the model (green line) is offset for clarity in all cases. {\bf g}, Simulated Ir-Ir partial xPDFs for undistorted cubic \fd3m\ (red line) and distorted tetragonal \i41amd\ (blue line) structures, with the associated difference between the tetragonal and cubic models (green line) underneath, offset for clarity. {\bf h}, Short range $1.5<r<6$~\AA\ fit of the distorted tetragonal model (red line) to the 500~K data (blue symbols), with associated difference curve underneath. Insets to {\bf a} and {\bf b} show evolution with reduced temperature of the cubic model fit residual, and the estimated local ODL distortion correlation length, respectively. Solid lines are guides to the eye.}
\end{figure*}

To explore the temperature dependence, the same analysis is carried out on PDFs measured at temperatures up to 780~K and representative fits are shown in Fig.~\ref{fig:twonm}(a-f).
The result of the fitting for the 500~K dataset is reproduced in Fig.~\ref{fig:twonm}(a) over the entire $r$-range, and then on a narrower $r$-scale in Fig.~\ref{fig:twonm}(c), with the residual signal highlighted.  The same signal in the residual is also evident at 232~K, 300~K and in the 780~K data (Fig.~\ref{fig:twonm}(d), (f) and (e), respectively). The 232~K dataset is from immediately above the MI transition temperature (226~K on warming).

To explore the structural origin of this residual signal we utilized structural models that allowed for distortions to the pyrochlore sublattice, and focused on a model in the \i41amd\ space group that was implicated in the early studies of the low T phase~\cite{furub;jpsj94}.
In particular, the tetragonal distortion lowers the symmetry of the regular Ir$_{4}$ pyrochlore tetrahedra, yielding 2 short and 4 long Ir-pair distances (Fig.~\ref{fig:onenm}(e)).
Comparison of PDFs computed from the distorted and undistorted models, as seen in Fig.~\ref{fig:twonm}(g), results in a difference curve that qualitatively reproduces the residual signal observed at 3.5~\AA, Fig.~\ref{fig:onenm}(b), when fitting with an undistorted cubic model.
However, the tetragonal distortion leads to additional features in the PDF which are not seen in the measured data, suggesting that the tetragonal distortion is not appropriate at all length scales.
Here we take advantage of the real-space nature of the PDF, and fit a tetragonal model over the narrow range $1.5<r<6$~\AA. This removes the residual signal at 3.5~\AA\ (Fig.~\ref{fig:twonm}(h)) and introduces only a single additional refinement parameter (tetragonal axis). The resulting fit produced a tetragonal distortion of 0.08(1)~\AA, which corresponds to long Ir-Ir bonds of 3.52~\AA\ and short Ir-Ir bonds of 3.44~\AA .

This disparity in symmetry paints a clear picture where Ir$_4$ tetrahedra, each with a local tetragonal distortion, are oriented in a disordered manner, such that individual distortions do not accumulate over long length scales, but rather average to a cubic symmetry. This is consistent with the observation that the low-$r$ portion of the PDF can be reproduced well only when using a tetragonally distorted model, but that this model fails to reproduce higher-$r$ features. Notably, the magnitude of the structural distortion at high temperature is $7\times$ smaller than the distortion corresponding to the dimer state.

The correlation length of the ordering of such distorted tetrahedra can in principle be extracted from the PDF.
Inspection of the residual curve in the 232~K data (Fig.~\ref{fig:twonm}(b)) suggests that at lower temperatures, though still above \tmit, the fit of the cubic model is worse than at higher temperature and that the residual signal extends over a wider-range of $r$, up to 40~\AA .
This would be the case if the short bonds were beginning to short-range order with some correlation length. The temperature dependence of the correlation length, $\xi$,  can be estimated using previously reported protocols~\cite{bozin;sr14} and further described in Methods Section. The result is shown in the inset to Fig.~\ref{fig:twonm}(b).
The correlation length is 6~\AA\ at high temperature, but smoothly increases to 20~\AA\ as the MI transition is approached.
This divergent behavior is mimicked if we consider the cubic model fit residual, $r_w$, as a function of temperature (inset to Fig.~\ref{fig:twonm}(a)).

\begin{figure*}[tb]
\includegraphics[width=1.0\textwidth]{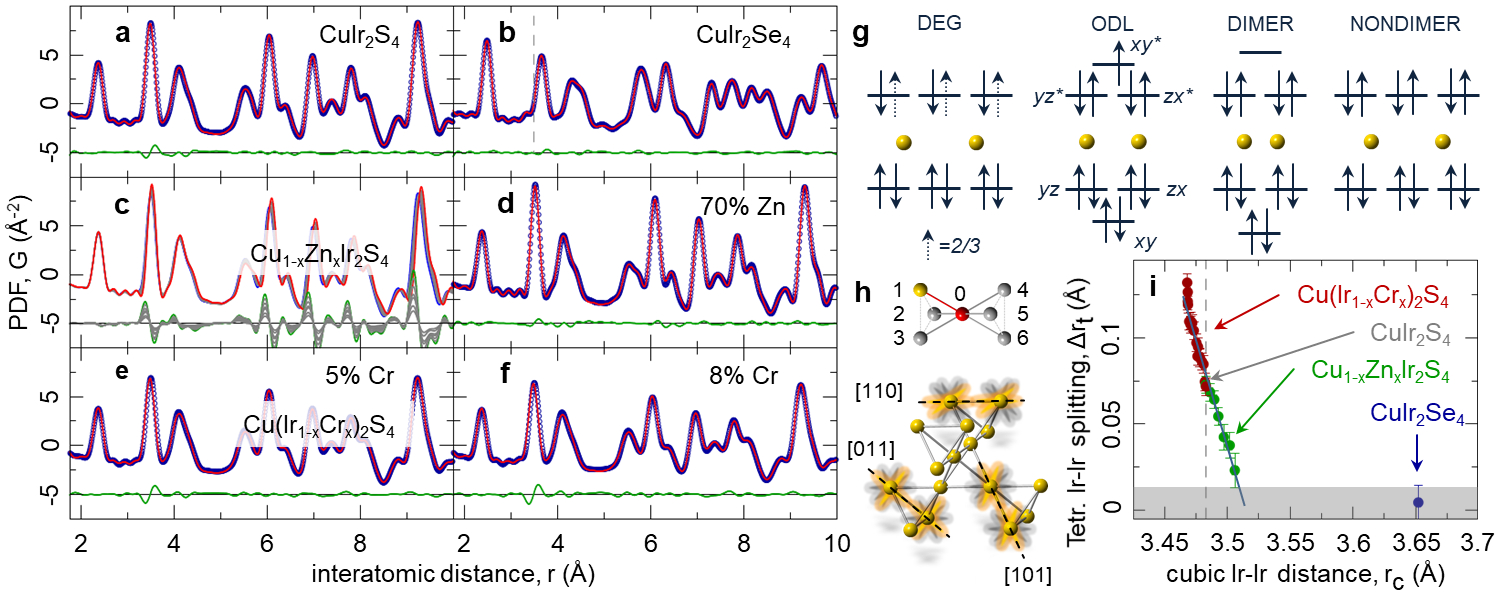}
\caption{\label{fig:threenm} {\bf Manipulation of the ODL state.} {\bf a}, Fit of the cubic model (line) to the 300~K \cis\ data (symbols). {\bf b}, Fit of the cubic model (line) to the 300~K \cise\ data (symbols). {\bf c}, Compositional stack of 300~K data for Zn-substituted \cis\ with Zn content ranging from 0~\% (blue) to 70~\% (red) in 10~\% increments (gray). The differences between the \cis\ parent and all other datasets are stacked underneath, offset for clarity. The largest difference between the 0~\% Zn and 70~\% Zn datasets is shown in green, other differences in gray, evolving uniformly with Zn content. {\bf d}, Fit of the cubic model (line) to the 300~K 70~\% Zn-substituted \cis\ data (symbols). {\bf e},{\bf f}, Fit of the undistorted cubic model (line) to the 300~K 5~\% and 8~\% Cr-substituted \cis\ data (symbols), respectively. {\bf g}, Molecular-orbital (MO) view, from left to right, of degenerate MO, degeneracy lifted MO, dimerized, and non-dimerized Ir-Ir contacts. In the legend, DEG (Ir$^{3.5+}$), ODL  (Ir$^{3.5+}$), DIMER  (Ir$^{4+}$), NONDIMER  (Ir$^{3+}$). {\bf h}, Sketch of [1,1,0]-type Ir \t2g\ overlaps (bottom) and six choices for each Ir to form an ODL state (top).
{\bf i},  Evolution of the ODL distortion, defined as the difference of the Ir-Ir nearest neighbor distance on a pyrochlore lattice extracted from local tetragonal model, with the average Ir-Ir separation in the cubic structure. These are extracted from fits to 300~K data of Cr-substituted (red, $0<x<0.6$) and Zn-substituted (green, $0<x<0.7$) samples, as well as pure CuIr$_2$S$_4$ (gray) and CuIr$_2$Se$_4$ (blue). Gray shaded region marks $2\sigma$ uncertainty for detecting small tetragonal distortions by the approach utilized here. In \cis\ there are 0.5 \t2g\ holes per Ir (one hole per pair).~\cite{khoms;prl05} Vertical gray dashed lines in {\bf b} and {\bf i} refer to \cis.}
\end{figure*}

The symmetry breaking implies a breaking of the degeneracy of orbitals~\cite{yagas;jpsj06}, which we refer to as an orbital degeneracy lifted state, dubbed ODL, on at least some of the Ir tetrahedra. The charge selects and preferrentially enters the lower energy orbitals, which may fluctuate among all the possible edges of the pyrochlore tetrahedra at high temperature (Fig.~\ref{fig:threenm}(h)).
Such a phenomenon could be caused by various driving forces, including Jahn-Teller effects, covalency, or spin-orbit coupling~\cite{strel;pnas16,strel;pu17}, and we do not speculate on the origin yet. However, we note that orbital selectivity impacts phenomena in diverse systems from VO$_{2}$~\cite{aetuk;np13}, K$_{2}$Cr$_{8}$O$_{16}$~\cite{bhobe;prx15}, and Sr$_{2-x}$Ca$_x$RuO$_{4}$~\cite{medic;prl14}, to FeSe~\cite{sprau;s17,kosti;nm18,yu;prl18}, but the persistence of orbital selectivity to such high temperatures in a disordered orbital liquid state has not been widely observed.  Our data show that the structural and metal-insulator transition on cooling corresponds not to the formation of an orbitally ordered state, but to the phase coherence and resulting long-range ordering of the pre-formed ODL objects.

%
\section{Electronic manipulation of the ODL state}
We now establish an electronic driving force for this ODL effect.
Iridium takes on a nominal $5d^{5.5}$ average valence configuration. In the cubic pyrochlore lattice the $t_{2g}$ orbitals are well separated in energy from the $e_g$ due to crystal field effects, and the $t_{2g}$ orbital of one Ir points directly towards the neighboring Ir ion~\cite{khoms;prl05,strel;pu17}.  The large extent of the 5$d$-states suggests a significant overlap of these orbitals and considerable covalency~\cite{strel;pu17}, though this is not required for the discussion.  We could then consider the orbital selectivity~\cite{khoms;b;tmc14}, to happen on a basis of molecular orbitals~\cite{lei;prb14} (inset to  Fig.~\ref{fig:onenm}(c)). In this case there would be a symmetry breaking into short and long edges on the pyrochlore tetrahedra depending on the electron filling of the molecular orbital but the incomplete filling of the $t_{2g}$ manifold provides a Jahn-Teller-like driving force for the distortion that lifts the orbital degeneracy. Each Ir has six neighbors to choose from and randomize orbital selectivity. These concepts are illustrated in Fig.~\ref{fig:threenm}(g) and~\ref{fig:threenm}(h) top.

Altering the charge state of the Ir ions offers a method by which we can test this hypothesis.
We have done this by doping Zn$^{2+}$ on the Cu$^{1+}$  site.
The zinc doping increases the electron count in the Ir sublattice, without
significantly disrupting the lattice~\cite{cao;prb01}.  The details are provided in Supplementary Information (SI) and summarized in Fig.~\ref{fig:threenm}.
Zn doping increases the electron count in an anti-bonding ODL state, marked with an asterisk in Fig.~\ref{fig:threenm}(g). If the observed structural distortion is driven by a local Jahn-Teller effect, the feature in the residual should diminish with increasing Zn content as doping electrons in an anti-bonding band destabilizes the ODL state. This is exactly what is observed (Fig.~\ref{fig:threenm}(c), (d)), establishing the electronic driving force for the effect.

We also consider the substitution of Chromium on the Iridium sublattice.  Chromium is a small ion and introduces a compressive chemical pressure.
It also introduces quenched defects into the Ir sublattice, disrupting the ability of Ir orbitals to order over long range at low temperature and suppressing the low-temperature orbital order state~\cite{bozin;sr14}.  Fits of the cubic model to two of the Cr doped data-sets are shown in Fig.~\ref{fig:threenm}(e) and (f).
They clearly show that the signal in the residual at 3.5~\AA\ remains robustly up to $x=8$\%, and indeed is stronger than in the \cis\ endmember, despite the absense of a symmetry-broken ground-state.  The compressive chemical pressure has the effect of stabilizing the ODL-state, possibly due to an increasing Ir-Ir \t2g\ orbital overlap, suggesting that the symmetry lowering is among molecular orbitals rather than atomic $d$-states.

Finally, we consider the sister compound \cise. In this case the ODL signature in the fit residual is absent, Fig.~\ref{fig:threenm}(b).
In \cise\ the electron counting arguments are the same as in \cis , with the Ir $t_{2g}$ states being at the Fermi-level, implying similar physics.
However, the Se ion is considerably larger than the S ion which results in larger inter-Ir distance and lesser tendency for Ir-Ir bond formation in the case of \cise.
In a traditional site-centered Jahn-Teller picture this would not affect the driving-force, indeed it may even make the Jahn-Teller distortion larger by lowering the elastic stiffness of the material.   However, if covalency between neighboring Ir ions is important, as we believe is the case here, the reduced orbital overlap of the Ir \t2g\ orbitals would reduce the splitting of the bonding and anti-bonding orbitals, which could work to quench the effect.
To test this idea, we plot the local tetragonality (from the tetragonal model fits to low-$r$ PDF), which is a measure of the symmetry breaking, as a function of the average Ir-Ir distance, $r_c$, from the cubic model for the same dataset (Fig.~\ref{fig:threenm}(i)).
Evidently, $r_c$ is much larger in \cise\ because of the large size of Se, but the tetragonality is zero: there is no local symmetry breaking observed in the \cise\ case.

The importance of covalency would imply that the orbital degeneracy lifting may be stabilized by pressure, since pressure would increase the overlap of neighboring \t2g\ orbitals. Such a stabilization is supported by transport measurements where an MIT was induced in \cise\ by pressure~\cite{furub;jpsj97}, suggesting it is a latent ODL system under ambient conditions. This result also provides an explanation of why the MIT temperature increases with pressure in \cis~\cite{ma;d-t17}, a trend opposite to that seen in conventional Fermi surface nesting driven charge-density-wave systems.

\section{Implications}
The observed fluctuating ODL state in \cis\ and its absence in \cise\ testify to the central importance of orbitals in this spinel. Since at high temperature all Ir are isovalent and each forms an ODL state with one of its six Ir-neighbors, charge disproportionation at MIT is a prerequisite for the dimerization (Fig.~\ref{fig:threenm}(g), (h)). While long range orbital order is not required for charge disproportionation and subsequent dimerization~\cite{bozin;sr14}, the ODL state evidently is. Destabilization of the ODL state by electron doping could drive the observed lattice expansion upon Zn substitution via local distortion release. Furthermore, we speculate that in \czis\ there may exist a marginal regime where the ODL fluctuations persist but do not freeze out at low temperature. Coincidentally, this may be in the Zn-concentration range where superconductivity is observed in this system~\cite{cao;prb01}. Importantly, given the $5d$ nature of the system where strong spin-orbit coupling is expected, our observations imply that significant \t2g-like orbital anisotropy is retained.

The characterization of the high temperature state of \cis\ as being an ODL state, made up of local symmetry broken objects stabilized by orbital degeneracy lifting, presents a possible unifying concept and a new lens through which to view multiple material systems.  Because the objects are local and fluctuating they are not observable in the crystal structure. Calculations that derive from the crystal structure, such as density functional theory calculations, should therefore be modified to account for the very different (tenths of an angstrom) bond-lengths that may be present in the material~\cite{varig;arxiv19}.  This would not be necessary if the ground state consisted simply of ODL objects whose orbitals become  ordered over long-range at low temperature.  However, often the ground state is quite different from this, as in \cis\ where it consists of charge-order, with structural and magnetic dimers, none of which persist above the MIT and into the ODL state~\cite{bozin;prl11}.    Likewise, if we view the insulating polaronic state in the colossal magnetoresistant 30\% doped La$_{1-x}$Ca$_x$MnO$_3$\ manganites~\cite{milli;n98} as an ODL state, the low-temperature ground-state has been shown to be absent structural distortions~\cite{billi;prl96} and is a Non-ODL state.  In the \cis\ case long-range ordering is presumably suppressed due to the geomteric frustration of disordering a short Ir-Ir bond over the six edges of the tetrahedron in the pyrochlore lattice, a problem that maps onto the Pauling ice rules~\cite{thygesen_orbital_2017,bramwell_spin_2001}.

The high-temperature ODL state is in general not just a disordered form of the ground-state and needs to be studied independently and in its own right.
%
%
This is not straightforward, requiring probes of local structure and the local electronic system.  The ODL objects are also, in general, fluctuating in the disordered ODL state thus requiring probes that are also faster than any fluctuation dynamics.  However, because the nature of the ODL formation is electronic, we expect that optically pumped ultrafast time-resolved measurements of local structure should be a powerful approach to investigate ODL~\cite{koch;arxiv19,konst;arxiv19}.

\section{Methods}
\label{sec:exp}
%
Polycrystalline \cis, \cise, \czis, and \cics\ samples were prepared following standard solid state routes in sealed, evacuated quartz ampoules. Stoichiometric quantities of the metals and elemental sulfur or selenium were thoroughly mixed, pelletized, and sealed under vacuum.
The ampoules were slowly heated to various temperatures in 650-1100~$^{o}$C range, as appropriate to targeted compositions, and held at these temperatures for several weeks with intermediate grinding
and pressing.  All products were found to be single phase based on laboratory x-ray powder diffraction. Standard characterization
of DC susceptibility and four-terminal resistivity of the samples were carried out in {\it Quantum Design} PPMS-9
and MPMS-XL5, and found to be in excellent agreement with other studies~\cite{furub;jpsj94,nagat;prb98,burko;prb00,cao;prb01,radae;n02,endoh;prb03}.

PDF data for \ranget\ were obtained using standard protocols~\cite{egami;b;utbp12} from synchrotron x-ray total scattering experiments carried out at the 28-ID-2 x-ray powder diffraction (XPD) beamline of the National Synchrotron Light Source II at Brookhaven National Laboratory. The setup utilized a 67.7~keV x-ray beam ($\lambda = 0.183$~\AA), a {\it Perkin Elmer} amorphous silicon detector, a closed cycle {\it Cryoindustries of America} helium refrigerator, and a gas flow reactor with flexible coil heater. Two dimensional (2D) diffraction data were collected in rapid acquisition mode,~\cite{chupa;jac03} with 60 s exposure time for each data set.
The raw 2D data (collected on warming) were integrated and converted to intensity versus $Q$ using the software Fit2D~\cite{hamme;hpr96}, where $Q$ is the magnitude of the scattering vector.
Data reduction and Sine Fourier transform of measured total scattering structure functions up to a momentum transfer of $Q_{max}= 25$~\RAA\ was carried out using the {\sc PDFgetX3}~\cite{juhas;jac13} program. PDF structure refinements and simulations were carried out using the {\sc PDFgui} program suite~\cite{farro;jpcm07}.

Correlation length estimate was based on a protocol utilizing a 4~\AA\ wide box car window integration of the residual difference between the data and the cubic \fd3m\ model. $\xi$ is then defined as the $r$ value at which the integral drops by a factor of 2 from its low-$r$ limit, with uncertainty of the estimate conservatively set to half of the window size, similar to correlation length estimates carried out in past PDF studies.~\cite{qiu;prl05,bozin;sr14}

%
%
%
\section{Acknowledgments}
Work at Brookhaven National Laboratory was supported by US DOE, Office of Science, Office of Basic Energy Sciences under contract DE-SC0012704. Work in the Materials Science Division of Argonne National Laboratory, was sponsored by the U.S. Department of Energy Office of Science, Basic Energy Sciences, Materials Science and Engineering Division. This research used 28-ID-2 beamline of the National Synchrotron Light Source II, a U.S. Department of Energy (DOE) Office of Science User Facility operated for the DOE Office of Science by Brookhaven National Laboratory.

\section{Author Contributions}
ESB, JFM and SJLB conceived and designed the research. CP, HCL, JFM, YSH and HZ developed and carried out the synthesis and did material characterizations. ESB and MA carried out PDF measurements and analysis. WGY provided theoretical inputs. ESB, SJLB and JFM wrote the manuscript with intellectual inputs and edits from all the authors.

\section{Competing Interests}
The authors declare no conflict of interest.

\section{Additional information}
Supplementary information is available in the online version of the paper. Reprints and permissions information is available online at www.nature.com/reprints. Correspondence and requests for materials should be addressed to ESB or SJLB.

\section{References}

\end{document}